\documentclass[a4paper]{article}

\usepackage{INTERSPEECH2021}
\usepackage{bbding} 
\usepackage{multicol}
\usepackage{multirow}
\usepackage{amsthm,amsmath,amssymb}
\usepackage{mathrsfs}
\usepackage{arydshln}

\title{Streaming non-autoregressive model for any-to-many voice conversion}
\name{Author Name$^1$, Co-author Name$^2$}
\name{Ziyi Chen$^{1,2}$, Haoran Miao$^{1,2}$, Pengyuan Zhang$^{1,2,*}$
\thanks{*corresponding author}}
\address{ $^1$Key Laboratory of Speech Acoustics \& Content Understanding, Institute of Acoustics, Chinese Academy of
Sciences, Beijing, China\\
	$^2$University of Chinese Academy of Sciences, Beijing, China}
\email{\{chenziyi, miaohaoran, zhangpengyuan\}@hccl.ioa.ac.cn}

\begin{document}

\maketitle
\begin{abstract}

\end{abstract}

Voice conversion models have developed for decades, and current mainstream research focuses on non-streaming voice conversion. However, streaming voice conversion is more suitable for practical application scenarios than non-streaming voice conversion. In this paper, we propose a streaming any-to-many voice conversion based on fully non-autoregressive model, which includes a streaming transformer based acoustic model and a streaming vocoder. Streaming transformer based acoustic model is composed of a pre-trained encoder from streaming end-to-end based automatic speech recognition model and a decoder modified on FastSpeech blocks. Streaming vocoder is designed for streaming task with pseudo quadrature mirror filter bank and causal convolution. Experimental results show that the proposed method achieves significant performance both in latency and conversion quality and can be real-time on CPU and GPU.


\section{Introduction}
Speaker identity is one of the important characteristics in human speech. In voice conversion, the speaker identity can be changed from one to another one while the linguistic information can be preserved \cite{sisman2020overview}. Traditional voice conversion focused on spectrum mapping using statistical and signal processing methods on parallel training data, which has the same linguistic content from different speaker, for example gaussian mixture model(GMM) \cite{toda2007voice, zen2008probabilistic}, vector quantization(VQ) \cite{abe1990voice}, fuzzy vector quantization \cite{shikano1991speaker}, frequency wrapping \cite{helander2008impact},  partial least square regression \cite{helander2010voice} and dynamic kernel partial least squares regression (DKPLS) \cite{helander2011voice}. 

With the development of deep learning methods in recent several years, voice conversion technology has rapid development and researches focus on from parallel data to non-parallel data voice conversion. There are several generative models for non-parallel voice conversion. CycleGAN-VC ~\cite{kaneko2018cyclegan, kaneko2019cyclegan} and StarGAN-VC ~\cite{kameoka2018stargan, kaneko2019stargan} use generator or conditional generator to convert source features into target features with adversarial training. Autoencoder based models, like AutoVC ~\cite{qian2019autovc}, VQVC ~\cite{wu2020one, chen21e_interspeech}, VAEVC ~\cite{kameoka2018acvae}, utilize bottleneck layers to disentangle speaker and linguistic information. There are also some phonetic posteriorgrams(PPGs) ~\cite{sun2016phonetic, zhou2019cross} and automatic speech recognition(ASR) ~\cite{liu2020transferring, zhang2020recognition} based voice conversion models, which take the advantage of pre-trained ASR model to achieve speaker-independent linguistic information. However, most of above researchs are usually about non-streaming voice conversion, which is limited in practical situation. And the research of streaming voice conversion is far less than that of non-streaming. Although the method proposed in ~\cite{saeki2020real} can be implemented in real time on CPU, it can only be used in intra-gender one-to-one voice conversion. FastS2S-VC ~\cite{kameoka2021fasts2s} can achieve real-time conversion on CPU and GPU but in many-to-many situation. Considering the application scenario of streaming voice conversion, the source speaker is usually unknown, so it is necessary to study streaming any-to-many voice conversion.

In this paper, we propose a streaming non-autoregressive any-to-many voice conversion model, which is a PPGs based voice conversion for achieving better conversion capability and stability. The proposed method includes a streaming transformer based acoustic model and a streaming vocoder modified from HiFi-GAN ~\cite{kong2020hifi}. The main contributions are as follows: (1) A real-time any-to-many voice conversion method on CPU and GPU is proposed. (2) Chunk based mask introduced in non-autoregressive acoustic model can alleviate the mismatch between training and inference for better conversion quality. (3) The proposed vocoder can achieve high quality in streaming situation while maintaining a fast inference speed.

The rest of the paper is organized as follows: Section 2 describes our proposed method thoroughly. Experimental configurations and results are shown in Section 3, while the conclusion is drawn in Section 4.

\section{Proposed Method}

This section describes the details of our proposed real-time voice conversion along with training procedures. Basically,  speech is transformed into PPGs by fixed streaming encode from pre-trained streaming hybrid CTC/Attention transformer based ASR model. Then PPGs are converted to mel spectrograms by streaming decoder. Finally mel spectrograms are converted to speech by streaming vocoder. 

\begin{figure*}[!htbp]
\begin{center}
\includegraphics[width=0.90\textwidth]{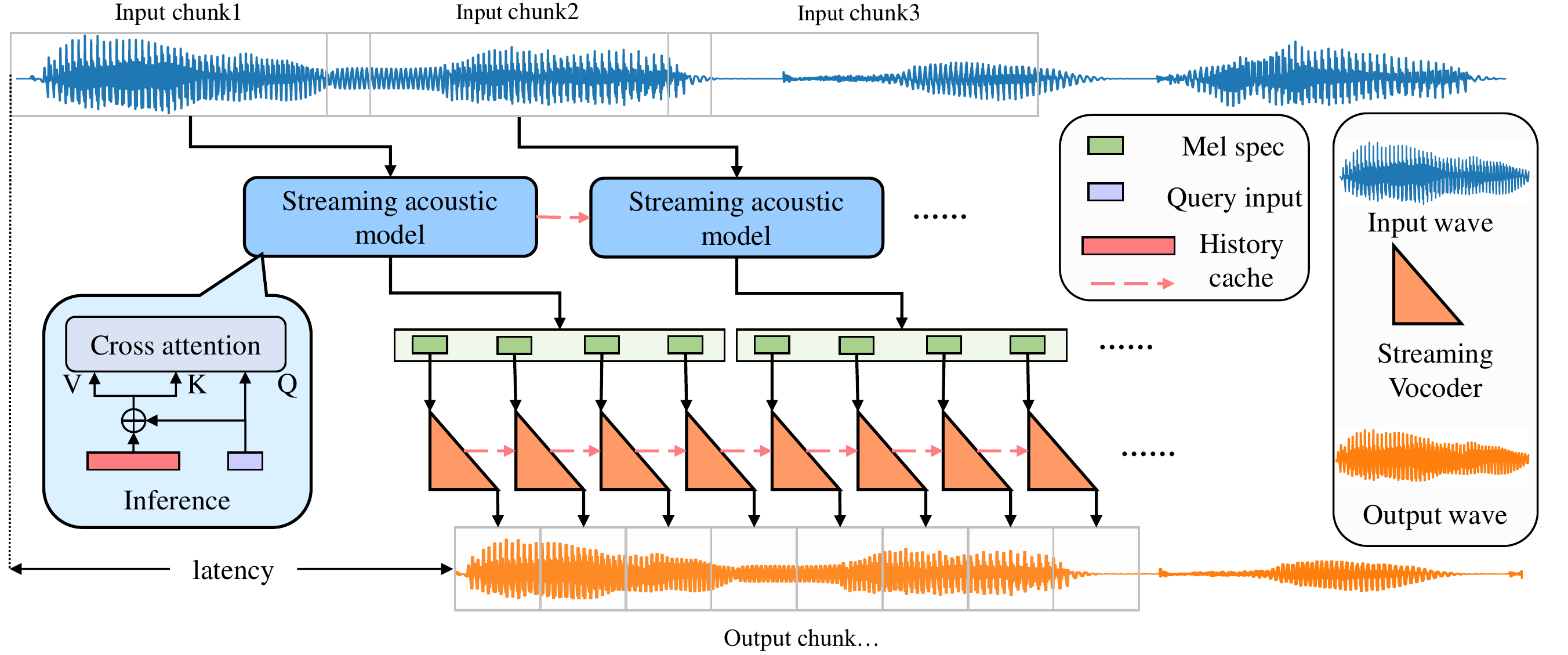}
\end{center}
\caption{Pipeline of proposed real-time voice conversion method}
\label{pipe}
\end{figure*}

Figure \ref{pipe} shows the overall pipeline of proposed method, streaming acoustic model and streaming vocoder are trained and inferred separately. And beyond that, streaming acoustic model is inferred chunk by chunk while streaming vocoder can be inferred frame by frame, which can reduce latency as much as possible. Streaming acoustic model and vocoder are elaborated in the following subsections.

\subsection{Streaming Acoustic Model}

Streaming acoustic model contains two parts: streaming encoder and streaming decoder, which are described in detail below.

\subsubsection{Streaming Encoder}

In vanilla transformer, queries, keys and values in self-attention layers are computed from whole speech, which is used in non-streaming tasks. But for streaming tasks, keys and values can be computed from accumulated speech while the query are computed from newly received chunks of speech. Although TransformerXL ~\cite{dai2019transformer} can be streaming, length of historical information is fixed and training process is slow. Therefore we consider dynamic historical speech for each query by using chunk based mask during training ~\cite{miao2022low}, which is similar with mask in Wenet ~\cite{zhang2020unified}.

\begin{equation}
    \begin{bmatrix}
1 & 0 & 0\\
1 & 1 & 0 \\
1 & 1 & 1 
\end{bmatrix} 
\otimes
\begin{bmatrix}
1 & 1 \\
1 & 1
\end{bmatrix} 
=
\left[
    \begin{array}{ll:ll:ll}
        1 & 1 & 0 & 0 & 0 & 0 \\
        1 & 1 & 0 & 0 & 0 & 0 \\
        \hdashline
        1 & 1 & 1 & 1 & 0 & 0 \\
        1 & 1 & 1 & 1 & 0 & 0 \\
        \hdashline
        1 & 1 & 1 & 1 & 1 & 1 \\
        1 & 1 & 1 & 1 & 1& 1
    \end{array}
\right]\label{eq1}
\end{equation}


Example form of chunk based mask is shown in Eq. \ref{eq1}, the connections between queries and keys are represented via boolean matrices, where 1 and 0 indicates connection and non-connection respectively. The first matrix defines the connections between chunks, the second one define the connection within chunks. The final connections can be represented by Kronecker product of above two matrices. This predefined matrix serves as chunk based mask in self-attention layers for parallel computing training. 

The chunk based mask is applied on the encoder of E2E ASR model. For capturing long-term information and short-term information in speech, dynamic chunk training is also applied on the encoder, which change uniformly in given range during training procedure. Then the pre-trained encoder on ASR data is used as the fixed encoder of proposed streaming acoustic model to extract PPGs. 

\subsubsection{Streaming Decoder}

Feed-Forward Transformer (FFT) blocks in FastSpeech~\cite{ren2019fastspeech} are adopted as the streaming decoder, which includes multi-head self-attention and 1D convolutions. Chunk based mask is also applied in self-attention layers of streaming decoder. To ensure that the output of current chunk is not affected by future chunk inputs, all vanilla convolutions in FFT are replaced by causal convolutions. 

\begin{figure}[!htbp]
\begin{center}
\includegraphics[width=0.40\textwidth]{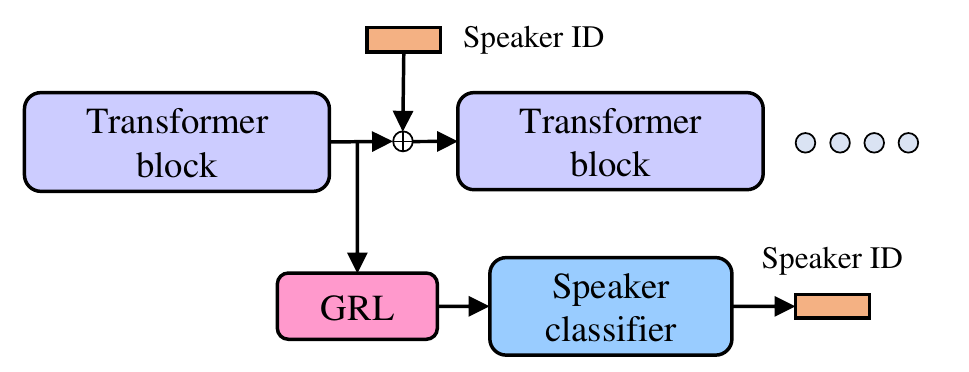}
\end{center}
\caption{Decoder of streaming acoustic model}
\label{decoder}
\end{figure}

In order to further remove residual speaker information in PPGs, a speaker adversarial network is applied after the first layer of decoder, which consists of a convolution based speaker classifier and a gradient reversal layer as in Figure \ref{decoder}.

\subsection{Streaming Vocoder}

\subsubsection{Multi-band HiFi-GAN}

In order to improve the inference speed while maintaining high generation quality, we propose a multi-band HiFi-GAN (MB HiFi-GAN) similar with multi-band MelGAN ~\cite{yang2021multi}. A stable and efficient low cost filter bank, called Pseudo Quadrature Mirror Filter Bank (PQMF), is employed for multi-band processing. The generator predicts all sub-bands simultaneously then they are synthesized together into final waveform by PQMF. To avoid the possible checkerboard artifacts caused by transpose convolution, we use temporal nearest interpolation layer followed by 1D convolution layer instead ~\cite{pons2021upsampling}. Times loss ~\cite{tian2020tfgan} and multi-resolution STFT loss ~\cite{yamamoto2020parallel} are added for stable training and alleviating central frequency band noise caused by PQMF.
 
\subsubsection{Smooth Method for Non-streaming Vocoder}

Due to receptive field of vanilla convolution in MB HiFi-GAN, output speech of current time frame is influenced by several or even more future frames. If mel spectrograms are sent chunk by chunk without overlap and then generated speech fragments are concatenated directly, there will be click sounds at the joint. We utilize hanning window with overlap generation to alleviate this problem. Hanning window is shown in Eq. \ref{eq3}, where $n$  belongs to $[-(N-1)/2, (N-1)/2]$ and $N$ denotes window length.

\begin{equation}
    \label{eq3}
    w[n] = \frac{1}{2}\Big[1+cos(2\pi*\frac{n}{N-1})\Big]
\end{equation}

For the $t$ th and $t+1$ th speech chunk of length $L$, window smoothness of overlap part is shown in Eq.~\ref{eq4}, where $i$ belongs to $[0, (N-1)/2]$ and $O$ denotes speech fragment after smoothness.

\begin{equation} \label{eq4}
    O = w[i] * C_{t}[L - \frac{(N-1)}{2} + i] + w[-i] * C_{t+1}[\frac{(N-1)}{2} - i]
\end{equation}

\subsubsection{Multi-band Streaming HiFi-GAN}

Although overlap with window based post-processing can alleviate clicking sounds in joint, the increase of overlapped length will improve computational cost and latency. To solve streaming problem further, we propose multi-band streaming HiFiGAN (MBS HiFi-GAN), which replace vanilla convolution with causal convolution. Through this method, output speech of current time frame will not be influenced by future frames, which means the same results in non-streaming and streaming generation.

\begin{figure}[!htbp]
\begin{center}
\includegraphics[width=0.45\textwidth]{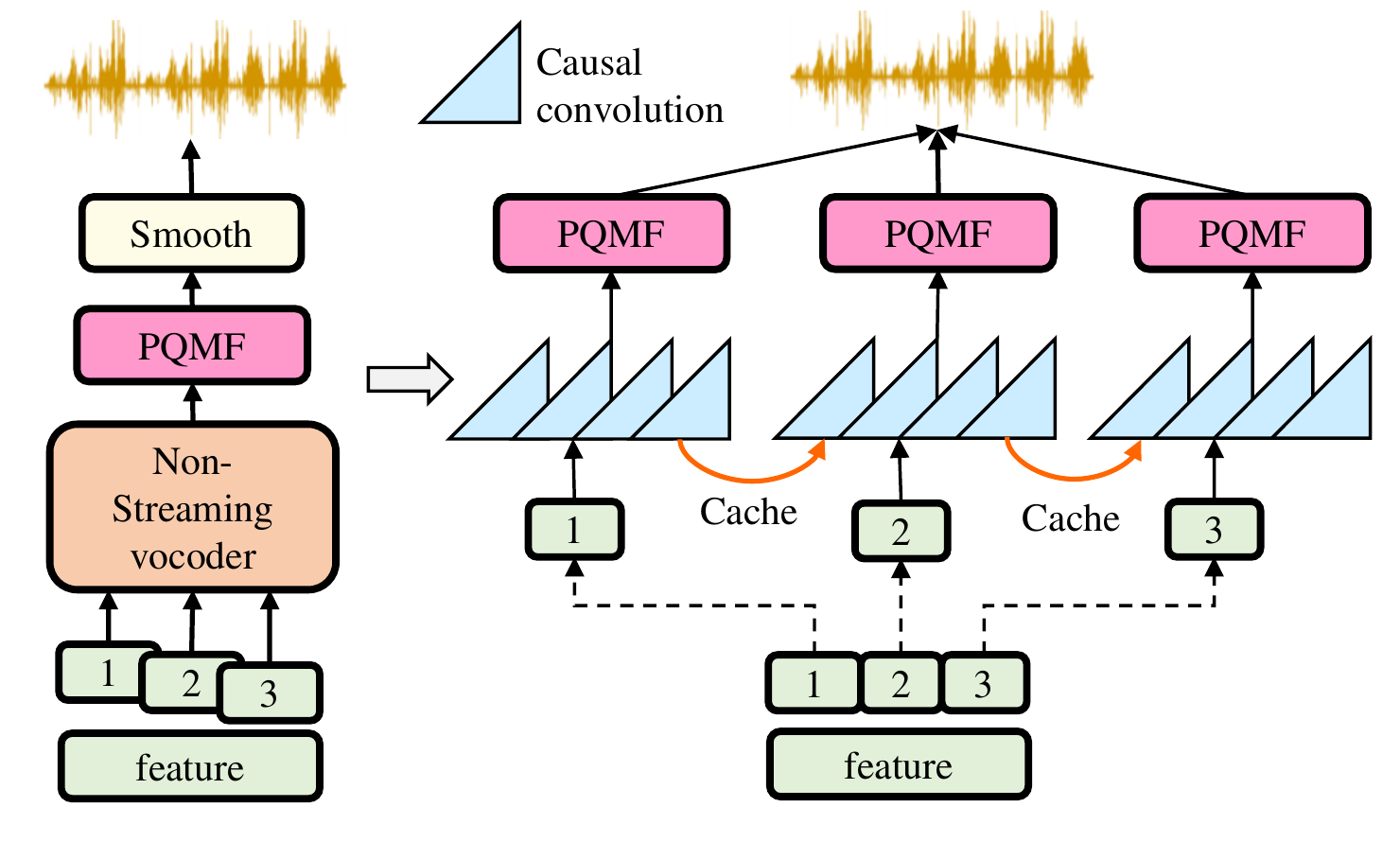}
\end{center}
\caption{Non-streaming and streaming vocoder}
\label{vocoder}
\end{figure}

The generation schematic diagram of MBS HiFi-GAN is shown in Figure ~\ref{vocoder}. it is observed that MBS HiFi-GAN almost don't introduce extra computational cost by saving the previous hidden states as caches in streaming situation. 

\subsection{Training Details}
ASR training loss $\mathcal{L}_{asr}$ combines CTC loss $\mathcal{L}_{ctc}$ and attention-based encoder-decoder (AED) loss $\mathcal{L}_{aed}$ as listed in Eq. \ref{asr}, where we set $\lambda=0.7$ .

\begin{equation}
    \label{asr}
    \mathcal{L}_{asr} = (1 - \lambda) \mathcal{L}_{ctc} + \lambda \mathcal{L}_{aed}
\end{equation}

Streaming acoustic model training loss $\mathcal{L}_{vc}$ is shown in Eq. \ref{acous_eq}, where $\mathcal{L}_{mel}$ denotes mel reconstruction loss and $\mathcal{L}_{spk}$ denotes speaker classifier loss.

\begin{equation}
    \label{acous_eq}
    \mathcal{L}_{vc} = \mathcal{L}_{mel} + \mathcal{L}_{spk}
\end{equation}
 
Generator training loss $\mathcal{L}_{sg}$ of streaming vocoder is shown in Eq. \ref{vocoder_eq}, where $\mathcal{L}_{g}$ denotes generator loss of original HiFi-GAN,  $\mathcal{L}_{time}$ denotes time loss, $\mathcal{L}_{stft}$ denotes multi-resolution loss, $\lambda_{time}=10.0$ and $\lambda_{stft}=2.0$.

\begin{equation}
    \label{vocoder_eq}
    \mathcal{L}_{sg} = \mathcal{L}_{g} + \lambda_{time}\mathcal{L}_{time} + \lambda_{stft}\mathcal{L}_{stft}
\end{equation}

\section{Experiements}

\subsection{Experimental Setup}
ASR models is trained on our own data about 1000 hours. The decoder of streaming acoustic model and vocoder are trained on open source mandarin datasets, including a 12 hours female dataset from databaker\footnote{https://www.data-baker.com/\#/data/index/source}, a parallel dataset of 5 hours male and 5 hours female audios from M2VoC challenge\footnote{http://challenge.ai.iqiyi.com/M2VoC}. All the audios are converted from 48KHz into 16KHz for experiments and evaluations.

40-dim MFCCs and 80-dim fbanks of 25ms frame length and 10ms frame stride are used for training ASR models and voice conversion models respectively. All the speed and latency evaluations are carried out on Intel(R) Xeon(R) Silver 4114 CPU and Nvidia Tesla P100 GPU without any extra engineering optimization. We set the number of threads to 1 by using \emph{torch.set\_num\_threads(1)} on CPU evaluations.

\subsection{Evaluations of Streaming Vocoder}

Experiments on proposed streaming vocoder are conducted for evaluating the performance. Mel cepstral distortion (MCD) objective evaluations of 800 selected items and speed evaluations are carried out. Results are shown in Table \ref{table1}. Although HiFi-GAN V1 achieves best performance in MCD but the inference speecd is much slower than other vocoders in CPU and GPU. HiFi-GAN V2 and V3 improve a lot in inference speed while get worst MCD. MB HiFi-GAN and MBS HiFi-GAN compromise inference speed and generation quality.

\begin{table}[!htbp]
\centering
\caption{MCD and speed evaluations of vocoders}
\begin{tabular}{lccc}
\toprule[1.0pt]
Model & MCD/dB  & CPU RTF & GPU RTF\\
\midrule
HiFi-GAN V1 & 3.029 & 0.902 & 0.0238 \\
HiFi-GAN V2 & 4.288 & 0.101 & 0.0096 \\
HiFi-GAN V3 & 4.290 & 0.102 & 0.0056 \\
MB HiFi-GAN & 3.235 & 0.190 & 0.0089\\
MBS HiFi-GAN     & 3.357 & 0.231 & 0.0085 \\
\bottomrule[1.0pt]
\end{tabular}
\label{table1}
\end{table}

\begin{figure}[!htbp]
\begin{center}
\includegraphics[width=0.45\textwidth]{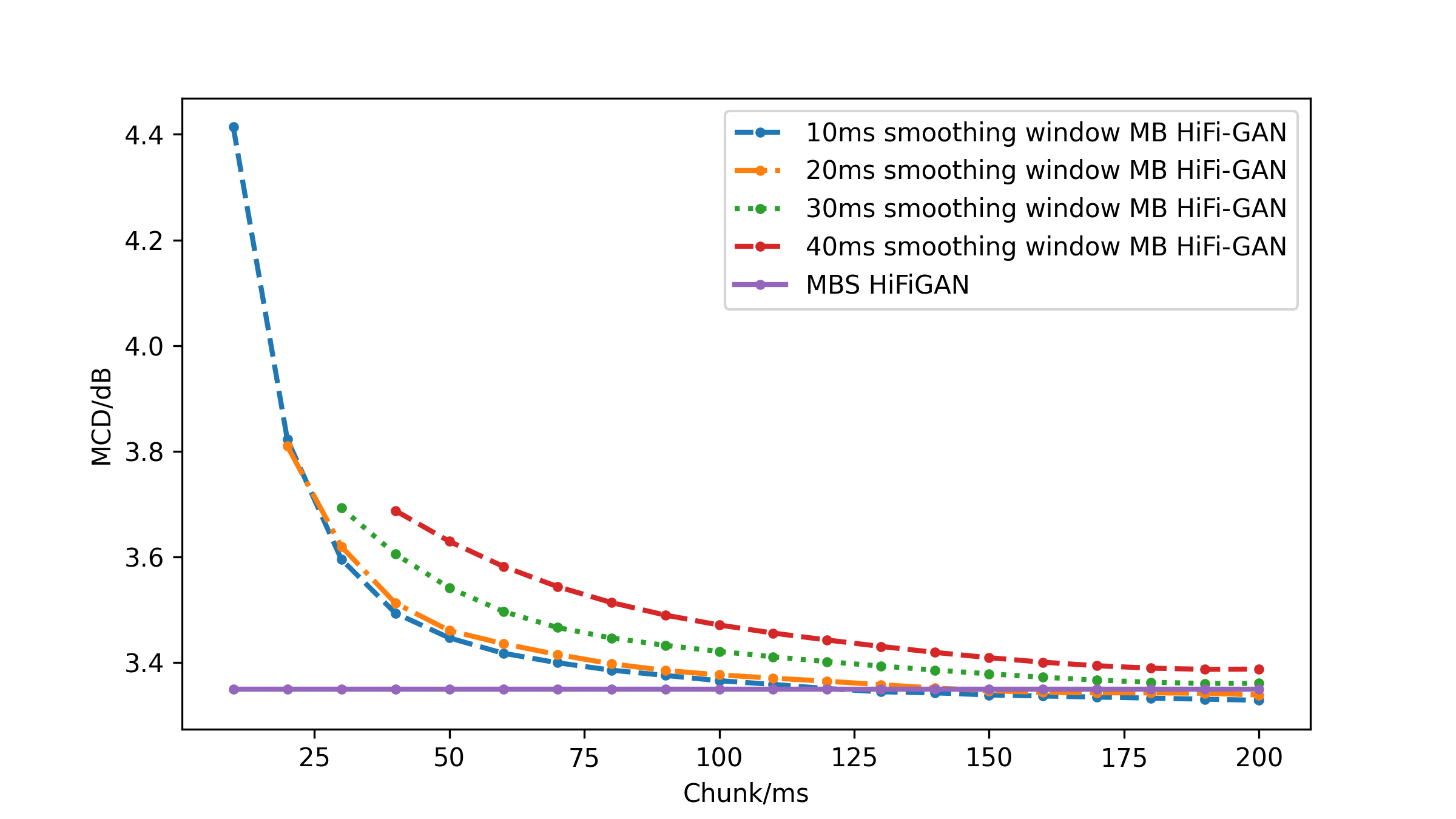}
\end{center}
\caption{MCD of streaming generation with MB HiFi-GAN and MBS HiFi-GAN in different chunk size}
\label{mcd1}
\end{figure}

In order to evaluate speech quality generated by different vocoders in streaming situation, MCD evaluations of fixed smooth window length with different chunk size are conducted. Results are shown in Figure \ref{mcd1}. As chunk size increase, MCD decreases while latency increases. Longer smoothing window can alleviate click sounds in the joint but results in MCD increase. The MCD of MBS HiFi-GAN does not change as the chunk size changes and is superior to other methods in small chunk size.

For low-latency tasks, speed experiments of MBS HiFi-GAN under small chunk size are carried out. Results are shown in Table \ref{table2}. In addition to 10ms chunk on CPU can not be real-time, the rest can be real-time.

\begin{table*}[!htbp]
\centering
\caption{Evaluations of proposed method with different chunk size}
\begin{tabular}{lccccccccccc}
\toprule[1.0pt]
\multicolumn{1}{c}{\multirow{2.5}{*}{Chunk Size}} &  \multicolumn{2}{c}{Encoder/ms} & \multicolumn{2}{c}{Decoder/ms} & \multicolumn{2}{c}{Vocoder/ms} & \multicolumn{2}{c}{Latency/ms} &  \multicolumn{2}{c}{Real-time} & \multirow{2.5}{*}{MCD/dB} \\ \cmidrule{2-11} 
\multicolumn{1}{c}{}                & CPU            & GPU           & CPU            & GPU           & CPU            & GPU           & CPU            & GPU   & CPU            & GPU     &    \\ \midrule
40ms                                                                   & 20.879         & 5.543         & 23.688         & 3.886         & 12.205         & 3.974         & 126.772        & 83.403   &      \XSolidBrush & \Checkmark & 6.688\\
80ms                                                                   & 24.900         & 5.575         & 24.979         & 3.897         & 12.205         & 3.974         & 172.084        & 123.446   &      \Checkmark & \Checkmark   & 6.475 \\
120ms                                                                  & 29.620         & 5.581         & 27.653         & 3.922         & 12.205         & 3.974         & 219.478        & 163.477  &      \Checkmark & \Checkmark   & 6.418 \\
160ms                                                                 & 40.117         & 5.602         & 30.321         & 3.943         & 12.205         & 3.974         & 272.643        & 203.519   &      \Checkmark & \Checkmark & 6.390 \\
200ms                                                                 & 44.571         & 5.807         & 43.548         & 4.162         & 12.205         & 3.974         & 330.324        & 243.942    &      \Checkmark & \Checkmark   & 6.382\\ 
\bottomrule[1.0pt]
\label{table7}
\end{tabular}
\end{table*}

\begin{table}[]
\centering
\caption{Inference speed of small chunk on CPU and GPU}
\begin{tabular}{lcccc}
\toprule[1.0pt]
Chunk size & 10ms & 20ms & 40ms & 80ms \\
\midrule
CPU/ms & 11.179 & 12.205 & 17.312 & 27.698\\
GPU/ms & 3.974 & 4.088 & 4.115 & 4.175\\
\bottomrule[1.0pt]
\end{tabular}
\label{table2}
\end{table}

\subsection{Evaluations of Streaming Acoustic Model}

To verify the effectiveness of the proposed method, objective evaluations of MCD, fundamental frequency root mean square error (F0-RMSE) and fundamental frequency correlation (F0-CORR) are conducted. We selecte 400 from M2VOC male and female dev set respectively for objective evaluation. Results are shown in Table \ref{table3}, where \emph{proposed} represents streaming acoustic model with speaker adversarial network, w/o \emph{sd} represents decoder is trained in non-streaming mode while inferred in streaming mode and w/o \emph{antispk} represents removing speaker adversarial network. All the models are inferred with 160ms chunk and 10 historical chunks and MBS HiFi-GAN vocoder. In Table \ref{table3}, \emph{proposed} achieves best performance among difference streaming acoustic models. The mismatch of non-streaming training and streaming inference cause the worst performance of w/o \emph{sd}. And w/o \emph{antispk} is worse than \emph{proposed}, which proves that speaker adversarial network is beneficial to better conversion capability. 

\begin{table}[!htbp]
    \centering
    \caption{Objective evaluations of different acoustic model}
    \begin{tabular}{lccc}
    \toprule[1.0pt]
    Model & MCD/dB & F0-RMSE/Hz & F0CORR\\
    \midrule
    \emph{proposed} & 6.390 & 43.024 & 0.554 \\
    - w/o $sd$ & 7.041 & 44.244 & 0.530 \\
    - w/o $antispk$ & 6.559 & 43.590 & 0.551 \\
    \bottomrule[1.0pt]
    \end{tabular}
    \label{table3}
\end{table}

\begin{table}[!htbp]
    \centering
    \caption{Subjective evaluations of different acoustic model}
    \begin{tabular}{lccc}
        \toprule[1.0pt]
        Model & Speech naturalness & Speaker similarity\\
        \midrule
        proposed & 3.67 $\pm$ 0.07  & 3.82 $\pm$ 0.06 \\
         - w/o $sd$ & 3.41 $\pm$ 0.04  & 3.35 $\pm$ 0.05 \\
         - w/o $antispk$ & 3.60 $\pm$ 0.05  & 3.61 $\pm$ 0.09 \\
        \bottomrule[1.0pt]
    \end{tabular}
    \label{table4}
\end{table}

Subjective evaluation on speech naturalness and speaker similarity of converted speech are conducted. Ten audiences are invited to give a 5-scale opinion score on both speaker similarity and naturalness. The subjective evaluation results are shown in Table ~\ref{table4}. It can be observed that there is a positive correlation between subjective and objective evaluations and proposed method achieves best performance in naturalness and speaker similarity. Generated audios can be found in demo page\footnote{https://miracyan.github.io/streamvc/}.

\subsection{Evaluations of Latency}

Latency is defined as the time from the start of source speech to the first packet of converted speech as in Figure \ref{pipe}. Due to the convolution downsampling layer, the proposed acoustic model can only deal with speech with integer multiples of 40 ms while proposed streaming vocoder can deal with speech with integer multiples of 10 ms. In Table \ref{table2}, proposed streaming vocoder can be real-time with 20ms chunk on CPU and 10ms chunk on GPU. Therefore, as long as it is ensured that the time for streaming acoustic model and vocoder to process the first packet of speech is less than input chunk time, real-time can be guaranteed. Because the convolution downsampling layer adopts non-padding convolution, it results in  the first packet needs to wait for another 30ms. Latency of proposed methods on CPU and GPU under different chunk size and MCD evaluations are shown in Table \ref{table7}. All model are trained and inferred with 10 historical chunks. It can be found that with the increase of chunk, the conversion quality of the model is also improved and proposed method can be real-time on CPU within 173ms lantecy and on GPU within 84ms latency. 


\section{Conclusions}

In this paper, we present a novel streaming any-to-many voice conversion method based on fully non-autoregressive models, which includes a streaming acoustic model and a streaming vocoder. Experiments on subjective, objective and latency evaluation demonstrate that significant performance of proposed method on streaming any-to-many voice conversion.

\bibliographystyle{IEEEtran}

\bibliography{mybib}


\end{document}